\documentclass[runningheads]{llncs}

\usepackage{amsmath}
\usepackage{epsfig,amssymb}
\usepackage{url}
\usepackage{epsfig}
\usepackage{tikz}

\newcommand{\semb}{[ \! [}
\newcommand{\seme}{] \! ]}
\newcommand{\comment}[1]{}

\def\norm#1{\mbox{$\| #1 \|_1$}}
\def\normd#1{\mbox{$\| #1 \|_2$}}
\def\norminfty#1{\mbox{$\| #1 \|_\infty$}}

\def\transpose#1{{}^t \! #1}

\def\stackreh#1#2{\mathrel{\mathop{#1}\limits_{#2}}}

\def\argmin#1#2{\mbox{argmin}_{|.|} (#1,#2)}

\def\forms{affine sets}
\def\pforms{perturbed affine sets}
\def\Forms{Affine sets}
\def\Pforms{Perturbed affine sets}
\def\form{affine set}
\def\pform{perturbed affine set}

\def\ua{\uparrow_\circ}
\def\da{\downarrow_\circ}
\def\ra{\rightarrow}
\def\bbr{{\Bbb R}}

\def\R{{\Bbb R}}

\newcommand {\N}{{\rm{I\!N}}}

 \newcommand\ForAuthors[1]
 {\par\smallskip                     
  \begin{center}
   \fbox
   {\parbox{0.9\linewidth}
    {\raggedright\sc--- #1}
   }
  \end{center}
  \par\smallskip                     %
 }

\newlength{\parskipLightVerbatim}
{\obeyspaces\gdef {\ }}
\def\TCode{\def\[{[}\def\]{]}\catcode`\[=1 \catcode`\]=2 \catcode`\{=12
\catcode`\}=12 \catcode`\&=12 \catcode`\#=12
\catcode`\~=12 \catcode`\_=12 \catcode`\^=12 \obeyspaces \tt}
 
\newenvironment{LightVerbatim}{%
  \begingroup \small \let\oldpar=\par \tolerance=10000 %
  \def\par{\mbox{}\oldpar\setlength{\leftskip}{\parskipLightVerbatim}\setlength{\parindent}{-\parskipLightVerbatim}} %
  \TCode \obeylines \parskip=0pt}
{\endgroup}

\begin{document}
\def\bbr{{\Bbb R}}
\def\ua{\uparrow_\circ}
\def\da{\downarrow_\circ}
\def\ra{\rightarrow}

\def\uan{\uparrow_n}
\def\dan{\downarrow_n}
\def\uam{\uparrow_-}
\def\dam{\downarrow_-}
\def\uap{\uparrow_+}
\def\dap{\downarrow_+}

\def\bfm#1{\protect{\makebox{\boldmath $#1$}}}
\def\a {\bfm{a}}
\def\b {\bfm{b}}
\def\c {\bfm{c}}
\def\d {\bfm{d}}
\def\e {\bfm{e}}
\def\f {\bfm{f}}
\def\g {\bfm{g}}
\def\h {\bfm{h}}
\def\ii {\bfm{i}}       
\def\j {\bfm{j}}
\def\k {\bfm{k}}
\def\l {\bfm{l}}
\def\m {\bfm{m}}
\def\n {\bfm{n}}
\def\o {\bfm{o}}
\def\p {\bfm{p}}
\def\q {\bfm{q}}
\def\r {\bfm{r}}
\def\s {\bfm{s}}
\def\t {\bfm{t}}
\def\u {\bfm{u}}
\def\vv {\bfm{v}}       
\def\w {\bfm{w}}
\def\x {\bfm{x}}
\def\y {\bfm{y}}
\def\z {\bfm{z}}
\def\B {\bfm{B}}
\def\C {\bfm{C}}
\def\DD{\bfm{D}}        
\def\E {\bfm{E}}
\def\G {\bfm{G}}
\def\H {\bfm{H}}
\def\J {\bfm{J}}
\def\L {\bfm{L}}
\def\M {\bfm{M}}
\def\N {\bfm{N}}
\def\O {\bfm{O}}
\def\P {\bfm{P}}
\def\Q {\bfm{Q}}
\def\S {\bfm{S}}
\def\T {\bfm{T}}
\def\U {\bfm{U}}
\def\V {\bfm{V}}
\def\W {\bfm{W}}
\def\X {\bfm{X}}
\def\Y {\bfm{Y}}
\def\Z {\bfm{Z}}

\pagestyle{headings}

\mainmatter

\title{
A Zonotopic Framework for Functional Abstractions}
\titlerunning{}

\author{Eric Goubault and Sylvie Putot}


\institute{CEA LIST, Laboratory for the Modelling and Analysis of Interacting Systems,\\
Point courrier 94, Gif-sur-Yvette, F-91191 France, \email{Firstname.Lastname@cea.fr}
}

\maketitle

\begin{abstract}
This article formalizes an abstraction of input/output relations, based on parameterized zonotopes,
which we call affine sets.
We describe the abstract transfer functions and prove their correctness, which allows 
the generation of accurate numerical invariants. Other applications range from compositional 
reasoning to proofs of user-defined complex invariants and test case generation. 
\end{abstract} 

\section{Introduction}
\label{intro}
We present in this paper an abstract domain based on affine arithmetic 
\cite{com-sto-93-aa} to bound the values of variables in numerical programs, with a real number semantics. 
Affine arithmetic can be conceived as describing particular
polytopes, called zonotopes \cite{ziegler}, which are bounded and
center-symmetric. But it does so by explicitly parametrizing the points, 
as affine combinations of symbolic variables, called noise symbols. 
This parametrization keeps, in an implicit manner,
the affine correlations between values of program variables, by sharing some
of these noise symbols. 
It is tempting then to attribute a meaning to these noise
symbols, so that the abstract elements we are considering are no longer
merely polytopes, but have a functional interpretation, due to their particular parametrization: 
we define abstract elements as tuples of affine forms, which we call \forms. 
They define a sound abstraction of relations that hold between the  current values
of the variables, for each control point, and the inputs of a program. 
The interests of abstracting input/output relations are well-known \cite{CousotCousot01-SSGRR}, we
mention but a few: more precise and scalable interprocedural abstractions,
proofs of complex invariants (involving relations between inputs and outputs), sensitivity analysis
and test case generation as exemplified in \cite{FMICS09}. 

An abstract domain relying on such affine forms has been described in \cite{CAVT09,DBLP:conf/sas/GoubaultP06,DBLP:journals/corr/abs-0807-2961}, 
but these descriptions miss complete formalization, and over-approximate the
input/output relations more than necessary. In this paper, we extend this preliminary work by 
presenting a natural framework for this domain, with a partial order relation that allows 
Kleene like iteration for accurately solving fixed point equations. In particular, a partial order that is now global to the abstract state,
and no longer defined independently on each variable, allows to use relations also between the special noise symbols created by
taking an upper bound of two affine forms. Our results are illustrated 
with sample computations and geometric interpretations.

A preliminary version of this abstract domain, extended to analyse the 
uncertainty due to floating-point computations, is used in practice in a real industrial-size static analyser
- FLUCTUAT -  whose applications have been described in \cite{FMICS09,FMICS07}. 
A preliminary version of this domain, dedicated to the analysis of computations in real numbers, is also 
implemented as an abstract domain - Taylor1+ \cite{CAVT09} - of the open-source library APRON \cite{mine:cav09}.

\paragraph{Related work}

Apart from the work of the authors already mentioned, that uses zonotopes in static analysis,
a large amount of work has been carried out mostly for reachability analysis in hybrid systems  
using zonotopes, see for instance \cite{DBLP:conf/hybrid/Girard05}. One common feature with our
work is the fact that zonotopic methods prove to be precise and fast. But in general, in hybrid systems
analysis, no union operator
is defined, whereas it is an essential feature of our work. Also, the methods
used are purely geometrical: no information is kept concerning input/output relationships, e.g. as
witnessed by the methods used for computing intersections \cite{1423050}.
Zonotopes have also been used in imaging, in collision detection for instance, see \cite{644241}, where
purely geometrical joins have been defined. 

Recent work in static analysis
by abstract interpretation for input/output relations abstraction and modular analyses can be found in
\cite{CousotCousot01-SSGRR}, where an example is given in particular using polyhedra. 
In \cite{CousotCousot93-1}, it is shown that some classical analyses (e.g. Mycroft's strictness analysis)
are input/output relational analyses (also called dependence-sensitive analyses).
Applications of abstractions of input/output relations have been developped, in particular
for points-to alias analysis, using summary functions, see for instance 
\cite{DBLP:conf/popl/ChatterjeeRL99}.

\paragraph{Contents}
In Section \ref{aa}, we quickly introduce the principles of affine arithmetic, and show the interest of a domain with explicit parametrization of zonotopes, compared to its geometric counterpart, through simple examples. Then in Section \ref{zonotopes}, we state properties 
of \forms. Introducing a matrix representation, we make the link between the 
\forms $\,$ and their zonotope concretisation. We then introduce \pforms, that will allow us to define a partially 
ordered structure. Starting with a thorough explanation of the intuition at Section \ref{introperturbed}, we then
 describe the partial order relation in Section \ref{order}, the monotonic abstract transfer functions
in Section \ref{abstracttransfer}, and the join operator 
in Section \ref{joinoperator}. For intrinsic reasons, our abstract domain does not have least upper bounds, but
minimal upper bounds. We show in Section \ref{completeness} that a form of bounded-completeness holds that allows 
Kleene-like iteration for solving fixed point equations. By lack of space, we do not demonstrate here the behaviour of 
our abstract domain on fixed-point computations, but results on preliminary versions of our domain are described 
in \cite{CAVT09,DBLP:journals/corr/abs-0807-2961}.

\section{Abstracting input/output relations with affine arithmetic}
\label{aa}

\paragraph{Affine arithmetic}
Affine arithmetic is an extension of interval arithmetic on affine forms, first introduced in \cite{com-sto-93-aa}, 
that takes into account affine correlations between variables. 
An {\em affine form} is a formal sum over a set of {\em noise symbols} 
$\varepsilon_i$ \[ \hat x \stackrel{def}{=} \alpha^x_0 + \sum_{i=1}^n \alpha^x_i \varepsilon_i,\]
with $\alpha^x_i \in \R$ for all $i$. 
Each noise symbol $\varepsilon_i$ stands for an independent component of the total
uncertainty on the quantity $\hat x$, its value is unknown but bounded in [-1,1]; 
the corresponding coefficient $\alpha^x_i$ is a known real value, which gives 
the magnitude of that component. The same noise symbol can be 
shared by several quantities, indicating correlations among them. These noise 
symbols can not only model uncertainty in data or parameters,
but also uncertainty coming from computation.

The semantics of affine operations is straightforward, non affine operations are linearized : we refer 
the reader to \cite{DBLP:conf/sas/GoubaultP06,DBLP:journals/corr/abs-0807-2961} for more details on the semantics
for static analysis.

\vspace{-.3cm}
\paragraph{Introductory examples}

\label{introex}

Consider the simple interprocedural program~:
\begin{center}
\begin{minipage}{5cm}
\begin{center}

\begin{LightVerbatim}
float main() {
  float x $\in$ \[-1,1\];
  return f(x)-x;
}
\end[LightVerbatim]
\end{center}
\end{minipage}
\begin{minipage}{5cm}
\begin{center}
\begin{verbatim}
float f(float x) {
  float y;
  if (x >= 0) y = x + 1;
  else y = x - 1;
  return y; }
\end{verbatim}
\end{center}
\end{minipage}
\end{center}
\vspace{-.1cm}

In order to analyse this program precisely, we need to infer the relation
between the input and output of function \texttt{f}, since the 
\texttt{main} function subtracts the input of \texttt{f} from its output.
We will show in Section \ref{introperturbed} that our method gives an
accurate representation of such input/output relations, at low cost,
easily proving here that \texttt{main} returns a number between -1 and 1. 
We will also show that even tight geometric representations of the image of \texttt{f} on \texttt{[a,b]}
may fail to prove this.

Another interest of our method is to allow compositional abstractions
for interprocedural calls \cite{CousotCousot01-SSGRR}, making our domain very scalable. 
For instance, the abstract value for the output of \texttt{f},
as found in Section \ref{introperturbed}, represents the fact that
its value is the value of the input plus an unknown value in [-1,1]. 
In fact a little more might be found out, which would lay the basis
for efficient disjunctive analyses, where we would find that the output of \texttt{f}
is its input plus an unknown value in $\{-1,1\}$. This is left
for future work. 
This compact representation can be used as an abstract {\em summary function} 
(akin to the ones of \cite{DBLP:conf/popl/ChatterjeeRL99} or of \cite{CousotCousot93-1})
for \texttt{f} which can then be reused without re-analysis for
each calls to \texttt{f}.
The complete discussion of this aspect is nevertheless outside the scope of this paper.

Last but not least, input/output relations that are dealt with by our
method allow proofs of complex invariants, and test case generation
at low cost. Consider for instance the following program, where \texttt{g} computes an approximation of the square root of \texttt{x}
using a Taylor expansion of degree 2, centered at point 1:

\begin{center}
\begin{minipage}{5cm}
\begin{center}
\begin{LightVerbatim}
float main() {
  float x $\in$ \[1,2\], z, t;
  z = g(x);
  t = z*z-x;
  return t; }
\end[LightVerbatim]

\end{center}
\end{minipage}
\begin{minipage}{5cm}
\begin{center}
\begin{verbatim}
float g(float x) {
  float y;
  y = 3/8.0+3/4.0*x-1/8.0*x*x
  return y;
}
\end{verbatim}
\end{center}
\end{minipage}
\end{center}

With our semantics,
we will find the following abstract value for \texttt{x}, \texttt{z} and
\texttt{t}:
$x  =  \frac{3}{2}+\frac{1}{2} \varepsilon_1$, 
$z = \frac{19}{16}+\frac{3}{16}\varepsilon_1-\frac{1}{64}\varepsilon_2$ and
$t = -\frac{567}{8192}-\frac{7}{128}\varepsilon_1-\frac{19}{512}\varepsilon_2
-\frac{169}{8192}\varepsilon_3$.
This proves that \texttt{z} is within $[\frac{63}{64},\frac{89}{64}]\sim
[0.984,1.391]$ (real result is $[1,1.375]$), and that
\texttt{t} is within $[-\frac{93}{512},\frac{329}{4196}]\sim [-0.182,0.078]$
(real result is $[-0.066,0]$). This means that we get
a rather precise estimate of the quality of the algorithm that approximates the square root.
Finally, examining the dependency of $t$ on the noise symbol modelling the input, we see 
that $\varepsilon_1=1$, that is $x=2$, is the most likely value for reaching the maximum of $t$, 
in absolute value. This input value is thus a good
test case to maximize the algorithmic error between the approximation of square root
and the real square root. Here it does indeed correspond to the worst case.
These applications are detailed in \cite{FMICS09}, and stronger statements
about test case generation can be found in \cite{SAS07}, where a generalized
form for abstract values is used for under-approximations.


\section{\Forms $\,$ and zonotopes : notations and properties \label{zonotopes}}

In what follows, we introduce matrix notations to handle tuples of affine forms, which we call 
\forms, and characterize the geometric concretisation of sets of values taken by these \forms.

We note ${\cal M}(n,p)$ the space of matrices with $n$ lines and $p$ columns of real coefficients.
An \form $\,$ expressing the set of values taken by $p$ variables over $n$ noise symbols 
$\varepsilon_i, \; 1 \leq i \leq n$, can be represented by a matrix $A \in {\cal M}(n+1,p)$. 

\begin{figure}
\vspace{-.5cm}
\begin{center}
\begin{tikzpicture}[scale=0.25] 
\filldraw[fill=yellow!50!white] (29,10) -- (29,12) -- (23,14) -- (19,14) -- (11,10) -- (11,8) -- (17,6) -- (21,6) -- (29,10); 
\node (C) [fill=black,inner sep=1pt,shape=circle] at (20,10)  {};
\draw[step=1.cm,gray,very thin] (10,5) grid (30,15); 
\draw[->] (10,5) -- (30,5) node[right]{$x$} ;
\draw[->] (10,5) -- (10,15) node[right]{$y$};
\foreach \x/\xtext in {10, 15, 20, 25, 30} 
\draw (\x cm,5 cm)  node[below] {$\xtext$};
\foreach \y/\ytext in {5, 10, 15} 
\draw (10 cm,\y cm) node[left] {$\ytext$};
\end{tikzpicture} 
\caption{Zonotope concretization $\gamma(A)$ of \form $\,$ \{(\ref{affex1})-(\ref{affex2})\} \label{fig_concret}}
\end{center}
\vspace{-1cm}
\end{figure}
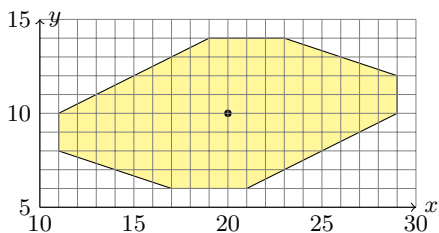

For example, consider the \form
\begin{eqnarray}
 \hat x &=& 20-4 \varepsilon_1 +2 \varepsilon_3 + 3 \varepsilon_4 \label{affex1} \\
 \hat y &=& 10-2\varepsilon_1 + \varepsilon_2 - \varepsilon_4, \label{affex2}
\end{eqnarray}
we have $n=4$, $p=2$ and~:
$\transpose{A}=\left(\begin{array}{ccccc}
20 & -4 & 0 & 2 & 3 \\
10 & -2 & 1 & 0 & -1
\end{array}\right)
$. 
Two matrix multiplications will be of interest in what follows~: 

\begin{itemize}
\item $A u$, where $u \in \R^p$, represents a linear combination of 
our $p$ variables, expressed on the $\varepsilon_i$ basis,
\item 
$\transpose{A} e$, where $e \in \R^{n+1}$, 
$e_0=1$ and $\norminfty{e}=\max_{0 \leq i\leq n} |e_i| \leq 1$,  
represents the vector of actual values that our $p$ variables take for the particular
values $e_i$ for each of our $\varepsilon_i$ noise variables. In this case, the additional symbol 
$\e_0$ which is equal to 1, accounts for constant terms, as done for instance in the zone abstract
domain \cite{zones}.
\end{itemize}
We formally define the zonotopic concretisation of \forms $\,$ by~: 
\begin{definition}\label{concretizations}
Let an \form $\,$ with $p$ variables over $n$ noise symbols, defined by a matrix $A \in {\cal M}(n+1,p)$. 
Its concretisation is the zonotope
$$\gamma(A)=\left\{\transpose{A} \transpose{\left( 1 | e  \right) }  \mid e \in R^{n}, \norminfty{e}\leq 1\right\} \subseteq \R^p. $$
We call its linear concretisation the zonotope centered on 0
$$\gamma_{lin}(A)=\left\{\transpose{A} e \mid e \in \R^{n+1}, \norminfty{e}\leq 1 \right\}\subseteq \R^p.  $$
\end{definition}
For example, Figure \ref{fig_concret} represents the concretization of the \form $\,$ defined by 
(\ref{affex1}) and (\ref{affex2}). It is a zonotope with center $(20,10)$ given by the vector 
of constant coefficients of the affine forms.

Zonotopes are particular bounded convex polyhedra \cite{ziegler}. A way
to characterize convex shapes is to consider support functions. 
For any direction $t
\in \R^p$, let $p_t$ the function which associates to all $x \in \R^p$,
$p_t(x)=\langle t,x\rangle$ where $\langle .,. \rangle$ is the
standard scalar product in $\R^p$, meaning that
$p_t(x)=\sum_{i=1}^p t_i x_i$. Level-sets of support functions, i.e.
sets defined by bounds on such functions characterize convex sets 
\cite{Bertsekas},
and nicely characterize zonotopes centered on 0:

\begin{lemma}\label{levelset}
Let $S$ be a convex shape in $\R^p$. Then $S$ can be characterized as the
(possibly infinite) intersection $\bigcap_{t \in \R^p} B_t$ of half-spaces of the form 
$$B_t = \{x \in \R^p \mid p_t(x)
\leq \sup_{y \in S} p_t(y)]\}$$

In case $S$ is a zonotope centered around $0$, it has finitely many
 faces with normals $t_i$ ($1 \leq i \leq k$), and this intersection
is finite:
$$S = \bigcap_{1 \leq i \leq k} \left\{x \in \R^p \mid  |p_{t_i}(x)| \leq \sup_{y \in S} p_{t_i}(y)\right\}$$
\end{lemma}

Furthermore, there is an easy way to characterize the linear concretization $\gamma_{lin}(A)$ (see also \cite{CAV09Girard}):
\begin{lemma}\label{supnorm}
Given a matrix $A \in {\cal M}(n+1,p)$, for all $t \in \R^p$, 
$\sup_{y \in \gamma_{lin}(A)} p_t(y)=\norm{A t}$, where
$\norm{e}=\sum_{i=0}^n |e_i|$ is the $\ell_1$ norm.
\end{lemma}

\noindent {\sc Proof.}
First of all, $\gamma_{lin}(A)$ is the image of the unit disc for the $L^{\infty}$ norm
by $\transpose{A}$ as we noted in Definition \ref{concretizations}. 
Therefore, $$\sup_{\{y \in \gamma_{lin}(A)\}} p_t(y)=
sup_{\{e \in \R^{n+1}, \norminfty{e}\leq 1\}} p_t(\transpose{A} e)$$
We now have 
$$\begin{array}{rcl}
p_t(\transpose{A} e) & = & \langle t, \transpose{A} e\rangle  = \langle A t, e\rangle 
 =  \sum_{i=0}^{n} \left(\sum_{j=1}^p a_{i,j} t_j\right) e_i \\
& \leq & \sum_{i=0}^n \left| \sum_{j=1}^p a_{i,j} t_j \right| 
\norminfty{e} = \norm{A t} \norminfty{e} 
\end{array}$$
This bound is reached for $e_i=sign\left(\sum_{j=1}^p a_{i,j} t_j\right)$, which is such that $\norminfty{e}=1$.\qed

\vskip .5cm

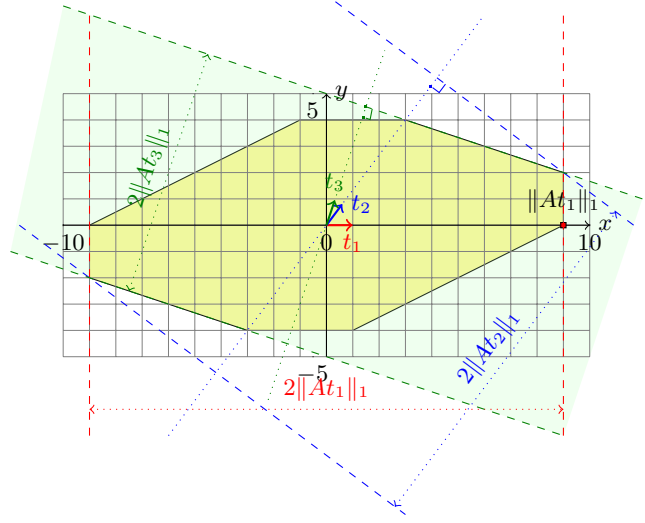
\begin{figure}
\vspace{-.5cm}
\begin{center}
\begin{tikzpicture}[scale=0.35] 
\usetikzlibrary{calc}
\filldraw[fill=yellow!50!white] (9,0) -- (9,2) -- (3,4) -- (-1,4) -- (-9,0) -- (-9,-2) -- (-3,-4) -- (1,-4) -- (9,0); 
\fill[green!30!white,opacity=0.2] (-10,25./3) -- (12,1) -- (9,-8) -- (-12,-1) -- (-10,25./3); 
\draw[step=1.cm,gray,very thin] (-10,-5) grid (10,5); 
\draw[->] (-10,0) -- (10,0)node[right]{$x$} ;
\draw[->] (0,-5) -- (0,5)node[right]{$y$};
\foreach \x/\xtext in {-10, 0/0, 10} 
\draw (\x cm,1pt) -- (\x cm,-1pt) node[anchor=north] {$\xtext$};
\foreach \y/\ytext in {-5, 5} 
\draw (-15pt,\y cm) -- (-15pt,\y cm) node[anchor=north] {$\ytext$};
\draw[red,thick,->] (0,0) -- (1,0) node[below]{$t_1$};
\node[shape=rectangle,draw,fill=red,inner sep=1pt,label=above:$$\norm{A t_1}= 9$$] (X1) at (9,0) {}; 
\draw[red,thin,dashed] (9,-8) -- (9,8);
\draw[red,thin,dashed] (-9,-8) -- (-9,8);
\draw[red,dotted,<->] (-9,-7) -- node[above] {$2 \norm{A t_1}$} (9,-7);
\draw[blue,thick,->] (0,0) -- (0.6,0.8) node[right] (t2) {$t_2$};
\coordinate (A2) at (-6,-8);
\coordinate (B2) at (6,8);
\draw[blue,dotted] (A2) -- (B2);
\coordinate (C2) at (35/3.0,0);
\coordinate (D2) at (0,35/4.);
\draw[blue,thin,dashed] (C2) -- (D2);
\node (G1)  at (intersection of C2--D2 and A2--B2) {}; 
\node (G2) [fill=blue,inner sep=0pt] at ($(G1)-(0.24,0.32)$)  {};
\node (t2b) at (0.8,-0.6)  {};
 \draw[blue,thin] (G2) -- ($(G2)+0.4*(t2b)$) -- ($(G1)+0.4*(t2b)$);
\draw[blue,thin,dashed] (-35/3.0,0) -- (3.0,-11.0);
\draw[blue,dotted,<->] (2.6,-10.7) -- node[sloped,above] {$2 \norm{A t_2}$} (11,0.5);
\draw[green!50!black,thick,->] (0,0) -- (2./6.325,6./6.325) node[above] (t3) {$t_3$};
\coordinate (A3) at (-7*2./6.325,-7*6./6.325);
\coordinate (B3) at (7*2./6.325,7*6./6.325);
\draw[green!50!black,dotted] (A3) -- (B3);
\coordinate (C3) at (-10,25./3);
\coordinate (D3) at (12,1);
\draw[green!50!black,thin,dashed] (C3) -- (D3);
\node (F1) [fill=green!50!black,inner sep=0pt] at (intersection of C3--D3 and A3--B3) {}; 
\node (F2) [fill=green!50!black,inner sep=0pt] at ($(F1)-0.25*(t3)$)  {};
\node (t3b) at (6./6.325,-2./6.325)  {};
 \draw[green!50!black,thin] (F2) -- ($(F2)+0.25*(t3b)$) -- ($(F1)+0.25*(t3b)$);
\draw[green!50!black,thin,dashed] (-12,-1) -- (9,-8);
\draw[green!50!black,dotted,<->] (-7.5,-2.5) -- node[sloped,above] {$2 \norm{A t_3}$} (-4.5,6.5);
\end{tikzpicture} 
\caption{Affine concretization $\gamma_{lin}(A')$ of \form $\,$ (\ref{affex1})-(\ref{affex2}) without its center \label{fig_concret_center}}
\end{center}
\vspace{-1cm}
\end{figure}

We illustrate Lemma \ref{supnorm} in Figure \ref{fig_concret_center}. Consider the matrix $A'$ associated to 
\form $\,$ \{(\ref{affex1})-(\ref{affex2})\} without its 
center. Its affine concretisation is the same zonotope as $\gamma(A)$ but centered on 0.
For $l \in R$, $t \in \R^p$, the $(l,t)$-level set corresponds to points 
on the hyperplane defined by : for $x \in \R^p$, $p_t(x)=\langle t,x \rangle = l$.
This hyperplane is orthogonal to the line $L_t$ going through 0, with direction
$t$. It intersects $L_t$ at a point
$y=\lambda t$ such that $\normd{t}^2 \lambda=l$. 
Given $t$ a direction in $\R^2$, the $(l,t)$-level set that intersects $\gamma_{lin}(A')$ with maximal value 
for $l$ realizes $l = \sup_{\gamma_{lin}(A')} p_t(y)=\norm{A' t}$ by Lemma \ref{supnorm}.
We now take three vectors $t$ such that $\normd{t}=1$. 
For $t_1=\transpose{(1,0)}$, $\norm{A' t_1}=9$, 
we find the maximum of its concretisation on the $x$-axis to be $9$. 
For $t_2 = \transpose{(3/5,4/5)}$, $\norm{A' t_2} =7/5$, and
$\gamma_{lin}(A') \subseteq H_{t_2}$, where $H_{t_2}$ is the region (or band)
between the line orthogonal to $t_2$ depicted as a blue dashed line and its symmetric with respect to zero.
For $t_3=\transpose{(2/\sqrt{40},6/\sqrt{40})}$ which is orthogonal to a face of the zonotope, $\norm{A' t_3} =3/4$ and $\gamma_{lin}(A') \subseteq
H_{t_3}$, which is the band between the two parallel faces in green. 

And indeed, for any matrix $A$, $\gamma_{lin}(A)$ is entirely 
described by providing the set of values $\norm{A t}$, where $t$ varies
among all directions in $\R^p$~: 
\begin{lemma}\label{utilzonotope1}
For matrices $X \in {\cal M} (n,p)$ and $Y \in {\cal M} (m,p)$, we have $\gamma_{lin}(X) \subseteq \gamma_{lin}(Y)$ if and only if $\norm{X u}\leq \norm{Y u}$ for all $u \in \R^p$.
\end{lemma}

\noindent {\sc Proof.}
Suppose first that $\norm{X u} \leq \norm{Y u}$ for
all $u \in \R^p$. 
~
By first part of Lemma \ref{levelset},
$$\gamma_{lin}(X)=\bigcap_{t \in \R^p} 
\{x \in \R^n \mid p_t(x)
\in [\inf_{y \in \gamma(X)} p_t(y), \sup_{y \in \gamma(X)} p_t(y)]\}$$
with $\sup_{y \in \gamma(X)} p_t(y) = -\inf_{y \in \gamma(X)} p_t(y)= \norm{X t}$ by Lemma \ref{supnorm}.
Thus
$$\gamma_{lin}(X) = \bigcap_{t \in \R^p} 
\{x \in \R^n \mid |p_t(x)| \leq \norm{X t} \} $$
$$\subseteq
 \bigcap_{t \in \R^p} 
\{x \in \R^n \mid |p_t(x)| \leq \norm{Y t} \} = \gamma_{lin}(Y).$$
\label{last}
Conversely, suppose $\gamma_{lin}(X)\subseteq \gamma_{lin}(Y)$. Then
$$\norm{X t} = \sup_{x \in \gamma_{lin}(X)} p_t(x) \leq \sup_{x \in \gamma_{lin}(Y)} p_t(x) = \norm{Y t}.$$\qed

\vskip .5cm

\section{\Pforms \label{pforms}}

\subsection{Rationale}

\label{introperturbed}

\begin{figure}
\vspace{-.5cm}
\begin{center}
\begin{tikzpicture}[xscale=2.,yscale=0.8] 
\filldraw[fill=red!20!white,draw=red!50!black,fill opacity=0.5] (-1,-2) -- (1,0) -- (1,2) -- (-1,0) -- (-1,-2); 
\filldraw[fill=blue!20!white,draw=blue!50!black,dashed,fill opacity=0.5] (-1,-2) -- (0,-1) -- (1,2) -- (0,1) -- (-1,-2); 
\coordinate [label=above:\textcolor{red!50!black}{$Z_1$}] (Z1) at (0.9,0); 
\coordinate [label=below:\textcolor{blue!50!black}{$Z_2$}] (Z2) at (0.1,1); 
\draw[very thick,black] (0,1) -- (1,2);
\draw[very thick,black] (-1,-2) -- (0,-1);
\draw[step=1.cm,gray,very thin] (-1,-2) grid (1,2); 
\draw[->] (-1.5,0) -- (1.5,0)node[below]{$x$} ;
\draw[->] (0,-2.5) -- (0,2.5)node[below,right]{$y$};
\foreach \x/\xtext in {-1, 0, 1} 
\draw (\x cm,1pt) -- (\x cm,-1pt) node[anchor=north] {$\xtext$};
\foreach \y/\ytext in {-2, -1, 1, 2} 
\draw (-0.15 cm,\y cm) -- (-0.15 cm,\y cm) node[anchor=north] {$\ytext$};
\end{tikzpicture} 
\caption{Two abstractions for the result of example function \texttt{f} defined Section \ref{aa}  \label{fig_exprog1}}
\end{center}
\vspace{-1cm}
\end{figure}
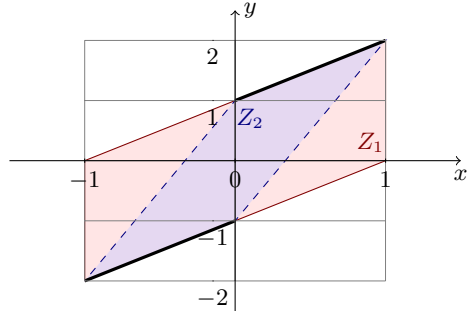

Let us get back to the program defining function {\tt f} in Section
\ref{aa}. We introduce a noise symbol $\varepsilon_1$ to
represent the range of values $[-1,1]$ for \texttt{x}. Using for example
the sub-optimal join operator described in Lemma \ref{nonoptimaljoin} to come, 
the \form $\,$ for \texttt{x} and \texttt{y} at the end of the program will be
$x = \varepsilon_1$, $y=\varepsilon_1+\eta_1$, with a new (perturbation) noise symbol $\eta_1$. 
The corresponding zonotope $Z_1$ is depicted in solid red in Figure \ref{fig_exprog1}. 

Now, a better {\em geometrical} abstraction of the
abstract value of \texttt{(x,y)} is the zonotope 
$Z_2$ depicted in dashed blue in Figure \ref{fig_exprog1}. Since
\texttt{y=x+1} for positive \texttt{x} and \texttt{y=x-1}
for negative \texttt{x}, we only have to include the two
segments in solid dark in the smallest zonotope as possible. This
is realized easily by a zonotope defined by the faces
$x-y \in [-1,1]$ and $y-3x \in [-3,3]$. Let us take a new symbol $\eta_2$
to represent $x-y$, and $\eta_3$ to represent $y-3x$. This
gives $x=-0.5\eta_2-0.5 \eta_3$ and $y=-1.5 \eta_2-0.5 \eta_3$. 
Although the corresponding blue zonotope $Z_2$ is strictly included
in the red zonotope $Z_1$, so it is {\em geometrically} more precise, 
we lose relations to the input values. Indeed, symbols $\varepsilon_i$ express 
dependencies to inputs of the program, whereas symbols $\eta_i$ do not.
 Thus, computing
\texttt{y} minus the input of \texttt{f}, as in the \texttt{main} function
of the example, gives $-\varepsilon_1
-1.5 \eta_2- 0.5 \eta_3 \in [-3,3]$. This range
is far less precise than using the representation $Z_1$, where we
find that this difference is equal to $\eta_1 \in [-1,1]$. \\

If we were not interested in input/output relations, a classical
abstraction based on \forms \; would be using
the geometrical ordering on zonotopes. We would say that \form \; $X$ is less or
equal than $Y$ iff $\gamma(X) \subseteq \gamma(Y)$. For the sake
of simplicity in the present discussion, suppose that $\gamma(X)$ and
$\gamma(Y)$ are centered on 0. 
By Lemma \ref{utilzonotope1}, we would then ask for  
$\norm{X t} \leq \norm{Y t}$ for all $t \in \R^p$.

Now, being interested in input/output relations, we will keep the  
existing symbols used to express possible ranges of values
of input variables (for instance, $\varepsilon_1$ defines the value of input  
variable \texttt{x} in the example above), and which should have a very strict  
interpretation, as well
as the noise symbols due to (non linear) arithmetic operations. We  
call them the
{\em central} noise symbols (such as $\varepsilon_1$).
And, to express  uncertainty on these relations due to possibly  
different execution paths, we will add additional noise symbols
which we call {\em perturbation} noise symbols (such as $\eta_1$ in the example above).

We now define an ordered structure using these two sets of noise symbols.

\subsection{Definition}
We thus consider {\em perturbed} affine sets $X$ as Minkowski sums 
\cite{Bertsekas} of 
a {\em central} zonotope $\gamma(C^X)$ and of a {\em perturbation}
zonotope (always centered on 0) $\gamma_{lin}(P^X)$~: 
\begin{definition}
We define a \pform \; $X$ by the pair of matrices \\ $(C^X, P^X) \in {\cal M}(n+1,p) \times {\cal M}(m,p)$. 
We call $C^X=(c^X_{ik})_{0\leq i \leq n, \; 1\leq k \leq p}$ the central matrix, and $P^X=(p^X_{jk})_{1\leq j \leq m, \; 1\leq k \leq p}$ 
the perturbation matrix. 

The perturbed affine form $\pi_k(X)=c^X_{0k}+\sum_{i=1}^n c^X_{ik} \varepsilon_i
+\sum_{j=1}^m p^X_{jk} \varepsilon^U_j$,
where the  $\varepsilon_i$ are the central noise symbols and the $\eta_j$ the perturbation or union 
noise symbols, describes the $k^{th}$ variable of X.
We call $\gamma(C^X)$ the central zonotope and $\gamma_{lin}(P^X)$ the perturbation zonotope. 
\end{definition}

For instance $Z_1$ as defined in Section \ref{introperturbed} is described by $C^1=(1 \mbox{ } 1)$, 
$P^1= (0 \mbox{ } 1)$ (first column corresponds to variable \texttt{x}, second
column, to \texttt{y}). $Z_2$ is described by 
$C^2 = (0 \mbox{ } 0)$ (the line corresponding to $\varepsilon_1$) and $P^2=\left(\begin{array}{cc}
-0.5 & -1.5 \\
-0.5 & -0.5
\end{array}\right)$ (the first line corresponds to perturbation symbol $\eta_2$, the second
to $\eta_3$).

\subsection{Ordered structure}
\label{order}
Expressing  $X$ less or equal than $Y$ on these \pforms $\,$ with the geometrical
order yields
$$\norm{C^X t}-\norm{C^Y t} \leq \norm{P^Y t}-\norm{P^X t}, \; \forall t \in \R^p.$$
But many transformations that leave $\norm{C^X t}$ and $\norm{C^Y t}$ fixed
for all $t$, and thus preserve that inequality, lose the intended meaning of the central noise symbols.
We can fix this easily, by strengthening this preorder.
Note that for all $t$, $\norm{C^X t}-\norm{C^Y t} \leq
\norm{(C^X-C^Y) t}$, so defining
$$X \leq Y  \mbox{ iff } 
\norm{(C^X -C^Y)t} \leq \norm{P^Y t} -\norm{P^X t}$$
should imply the geometrical ordering at least (as we will prove
in Lemma \ref{orderimpliesinclusion}). The good point is that
no transformation on the central noise symbols is allowed any
longer using this preorder (as the characterization of the
equivalence relation generated by this preorder will show, see Lemma
\ref{preorder-lemma}), keeping a strict interpretation of the noise symbols
describing the values of the input variables, hence the input/output 
relations.

We now formalize and study this stronger order: 
\begin{definition}\label{preorder-def}
Let $X=(C^X, P^X)$, $Y=(C^Y, P^Y)$ be two \pforms $\,$ in ${\cal M}(n+1,p) \times {\cal M}(m,p)$. 
We say that $X\leq Y$ iff 
\[ \sup_{u \in \R^p} \left(\norm{(C^Y-C^X)u}+\norm{P^Xu}-\norm{P^Yu}\right) \leq 0   \]
\end{definition}
Coming back to our example of Section \ref{introperturbed}, $\gamma(Z_2) \subseteq \gamma(Z_1)$ but 
$Z_2 \not \leq Z_1$. 
Take for instance $t=\transpose{(1, 1)}$. Then
$\norm{(C^1-C^2) t}+\norm{P^2 t}-\norm{P^1 t}=2+3-1=4>0$.

\begin{lemma}\label{preorder-lemma}
The binary relation $\leq$ of Definition \ref{preorder-def} is a preorder.
The equivalence relation generated by this preorder is
$X \sim Y$ iff by definition $X \leq Y$ and $Y \leq X$. It can be characterized
by $C^X=C^Y$ and $\gamma_{lin}(P^X)=\gamma_{lin}(P^Y)$ (geometrically speaking, as sets). 
We still denote $\leq/\sim$ by $\leq$ in the rest of the text.
\end{lemma}
\noindent {\sc Proof.}
Reflexivity of $\leq$ is immediate. Suppose now $X\leq Y$ and $Y \leq Z$, then for all $u \in
\R^p$:
$$\begin{array}{rcl}
\norm{(C^Y-C^X)u} & \leq & \norm{P^Y u}-\norm{P^X u} \\
\norm{(C^Z-C^Y)u} & \leq & \norm{P^Z u}-\norm{P^Y u} \\
\end{array}$$
Using the triangular inequality, we get 
$$\begin{array}{rcl}
\norm{(C^Z-C^X)u} & \leq & \norm{(C^Z-C^Y)u}+
\norm{(C^Y-C^X)u} \\
& \leq & \norm{P^Z u}-\norm{P^Y u}+\norm{P^Y u}-\norm{P^X u} \\
& \leq & \norm{P^Z u}-\norm{P^X u}
\end{array}$$
implying $X \leq Z$, hence transitivity of $\leq$.

Finally, $X \leq Y$ and $Y \leq X$ imply that for all $u \in \R^p$, 
$\norm{(C^Y-C^X)u}$ is less or equal than $\norm{P^Yu}-\norm{P^Xu}$ and is also 
less or equal than $\norm{P^Xu}-\norm{P^Yu}$. Hence $(C^Y-C^X)u=0$ for all $u$, meaning
$C^Y=C^X$ and $\norm{P^Xu}=\norm{P^Yu}$ for all $u$. 
By Lemma \ref{utilzonotope1} this exactly means that $\gamma(P^X)=\gamma(P^Y)$.\qed 
\vskip .5cm

\begin{lemma}\label{orderimpliesinclusion}
Take $X=(C^X, P^X)$ and $Y=(C^Y, P^Y)$. Then $X \leq Y$ implies
$$\gamma\left(\begin{array}{c}
C^X \\
\hline
P^X
\end{array}\right) \subseteq
\gamma\left(\begin{array}{c}
C^Y \\
\hline
P^Y
\end{array}\right)$$
or said in a different manner: $\gamma(C^X)\oplus\gamma_{lin}(P^X) \subseteq \gamma(C^Y) \oplus
\gamma_{lin}(P^Y)$ where $\oplus$ denotes the Minkowski sum. 
Note that $X \leq Y$ implies $\gamma_{lin}(P^X)\subseteq \gamma_{lin}(P^Y)$. 
\end{lemma}

\noindent {\sc Proof.}
It is easy to prove that 
$\gamma_{lin}\left(\begin{array}{c}
C^X \\
\hline
P^X
\end{array}\right) \subseteq
\gamma_{lin}\left(\begin{array}{c}
C^Y \\
\hline
P^Y
\end{array}\right)$ given that $X \leq Y$, using Lemma \ref{utilzonotope1} and the triangular
inequality for $\norm{.}$.

However, what we want is a little stronger. In order to derive it, we define, 
for all matrix $A$ of dimension $(n+1)\times p$, a matrix $\widetilde{A}$ of dimension $(n+1)\times (p+1)$ by
$$\widetilde{A}=\left(
\begin{array}{c|c}
\begin{array}{c} 
1 \\
0 \\
\vdots \\
0 
\end{array}
& A
\end{array}
\right)$$
The interest of this transformation, 
is that the zonotopic concretisation $\gamma(A)$ is a particular
face (which is the intersection with an hyperplane) of the 0-centered zonotope $\gamma_{lin}(\widetilde{A})$~: 
\begin{equation}
\label{conclin}
\gamma(A)=\gamma_{lin}(\widetilde{A}) \cap \{(1,x_1,\ldots,x_p) \mid (x_1,\ldots,x_p) \in \R^p  \}.
\end{equation}

 We now prove 
$\gamma_{lin}\widetilde{\left({\begin{array}{c}
C^X \\
\hline
P^X
\end{array}}\right)}
\subseteq
\gamma_{lin}\widetilde{\left({\begin{array}{c}
C^X \\
\hline
P^X
\end{array}}\right)}.$
For all $t=\transpose{(t_0,\ldots,t_p)} \in \R^{p+1}$,
$
\norm{
\widetilde{\left({\begin{array}{c}
C^X \\
\hline
P^X
\end{array}}\right)} t}
- 
\norm{\widetilde{\left({\begin{array}{c}
C^Y \\
\hline
P^Y
\end{array}}\right)} t}
$
$$\begin{array}{rcl} 
 & = & \norm{\widetilde{C^X} t}-\norm{\widetilde{C^Y} t}
+\norm{P^X t}
-\norm{P^Y t} \\
& = & 
\mid t_0+\sum_{k=1}^p c^X_{0,k} t_k \mid
- \mid t_0+\sum_{k=1}^p c^Y_{0,k} t_k \mid
+\norm{(c^X_{i,k})_{1 \leq i \leq n, 1 \leq k \leq p} \transpose{(t_1,\ldots
t_p)}}\\
& & 
-
\norm{(c^Y_{i,k})_{1 \leq i \leq n, 1 \leq k \leq p} \transpose{(t_1,\ldots
t_p)}}
+\norm{P^X t}
-\norm{P^Y t} \\
& \leq &
\mid \sum_{k=1}^p c^X_{0,k} t_k 
- \sum_{k=1}^p c^Y_{0,k} t_k \mid
+\norm{(c^X_{i,k})_{1 \leq i \leq n, 1 \leq k \leq p} \transpose{(t_1,\ldots
t_p)}}
\\
& & 
-
\norm{(c^Y_{i,k})_{1 \leq i \leq n, 1 \leq k \leq p} \transpose{(t_1,\ldots
t_p)}}
+\norm{P^X t}
-\norm{P^Y t} \\
& \leq &
\norm{(C^Y - C^X)t}
+\norm{P^X t}
-\norm{P^Y t} 
 \leq  0 
\end{array}$$
Hence by Lemma \ref{utilzonotope1},
$\gamma_{lin}\widetilde{\left(\begin{array}{c}
C^X \\
\hline
P^X
\end{array}\right)}
\subseteq
\gamma_{lin}\widetilde{\left(\begin{array}{c}
C^X \\
\hline
P^X
\end{array}\right)}$
which, by (\ref{conclin}), implies the result.
\qed

\vskip .5cm

The order we define is in fact essentially more complex than the inclusion ordering, while
still being computable:

\begin{lemma} \label{decidable}
The partial order $\leq$ is decidable, with a complexity bounded by a polynomial in $p$ and an
exponential in $n+m$.
\end{lemma}

\noindent {\sc Proof.}
The problem can be solved using $O(2^{(n+m)})$ linear programs. Let $X=(C^X, P^X)$, $Y=(C^Y, P^Y)$ be two \pforms $\,$ in ${\cal M}(n+1,p) \times {\cal M}(m,p)$. 
We want to decide algorithmically whether $X\leq Y$ that is 
\[ \sup_{u \in \R^p} \left(\norm{(C^Y-C^X)u}+\norm{P^Xu}-\norm{P^Yu}\right) \leq 0   \]

Looking at the proof of Lemma \ref{supnorm}, we
see that $$\norm{A u}=
\sup_{\{e \in \R^{n+1}, \norminfty{e}\leq 1\}} 
\sum_{i=0}^{n} \left(\sum_{j=1}^p a_{i,j} u_j\right) e_i$$
and that this bound is reached for $e \in \R^{n+1}$ such that 
for all $i$, $e_i=1$ or $e_i=-1$.

We therefore produce, for each $e \in \R^{n+1}$, $f \in \R^{m+1}$ and 
$g \in \R^{m+1}$, 
with, for all $i$, $e_i=1$ or $e_i=-1$, $f_i=1$ or $f_i=-1$ , $g_i=1$ or $g_i=-1$,  
the
following linear program:
$$sup_{u \in \R^p} \left(\sum_{i=0}^n \sum_{j=1}^p (c^Y_{i,j} -c^X_{i,j} ) e_i u_j  
+\sum_{i=1}^m \sum_{j=1}^p p^X_{i,j} f_i u_j
-  \sum_{i=1}^m \sum_{j=1}^p p^Y_{i,j} g_i u_j\right)
$$
subject to 
\begin{eqnarray*}
\left(\sum_{j=1}^p (c^Y_{i,j} -c^X_{i,j} ) u_j \right) e_i &\geq & 0, \; \forall 0 \leq i \leq n \\
\left(\sum_{j=1}^p p^X_{i,j} u_j \right) f_i &\geq & 0, \; \forall 1 \leq i \leq n \\
\left(\sum_{j=1}^p p^Y_{i,j} u_j \right) g_i &\geq & 0, \; \forall 1 \leq i \leq n 
\end{eqnarray*}

that we solve using any linear program solver (with polynomial complexity). We then check for each problem that 
it is either not satisfiable or its supremum is negative or zero.\qed

\vskip .5cm

Hopefully, there is no need to use this general decision procedure
in a static analyser by abstract interpretation. We refer the reader to the end
of Section \ref{completeness} for a discussion on this point.

\subsection{Extension of affine arithmetic on perturbed affine forms}

\label{abstracttransfer}

\subsubsection{Interpretation of assignments and correctness issues}

We detail below the interpretation of arithmetic expressions, dealing first with
affine assignments, that do not lose any precision. 
We use a very simple
form for the multiplication. There are in fact more precise ways to compute assignments
containing polynomial expressions. Firstly, the multiplication formula can be improved, see
\cite{CAVT09,DBLP:conf/sas/GoubaultP06}. Secondly, when interpreting a non-linear assignment, 
it is better in practice to introduce new noise symbols for the entire expression, and not for every 
non linear elementary operation as we present here. But for sake of simplicity, we do not describe this here. 
Note also that we would need formally to prove that projections onto a subset of variables (change of scope),
and renumbering of variables are monotonic operations, but these are easy checks and we omit them here.
Note finally that the proofs of monotonicity of our transfer functions are not only convenient for
getting fixpoints for our abstract semantics functionals. They are also necessary for proving the correctness of our
approach. As already stated in \cite{DBLP:conf/sas/GoubaultP06,DBLP:journals/corr/abs-0807-2961}, 
the correctness criterion we need relies on the property that whenever 
$X \leq Y$ are two \pforms, 
all future evaluations using expressions $e$ give smaller concretisations starting with $X$ than
starting with $Y$, i.e. $\gamma(\semb e \seme X) \subseteq \gamma(\semb e \seme Y)$. This
is proven easily as follows: as $\semb e \seme$ is a composite of monotonic
functions, $\semb e \seme X \leq \semb e \seme Y$. The conclusion holds because of Lemma
\ref{orderimpliesinclusion}.

\subsubsection{Affine assignments}

\label{affineassign}

We first define the assignment of a possibly unknown 
constant within bounds $a,b \in \R$ to a (new) variable,
$x_{p+1}:=[a,b]$:

\begin{definition}
Let $X=(C^X,P^X)$ be a \pform $\,$ in ${\cal M}(n+1,p) \times {\cal M}(m,p)$ 
and $a, b\in \R$. We define $Z= \semb x_{p+1}=[a,b] \seme X \in {\cal M}(n+2,p+1) \times {\cal M}(m,p+1)$ 
with~:
\begin{itemize}
\item $c^Z_{i,k}=c^X_{i,k}$ for all $i=0,\ldots,n$, 
$k=1, \ldots, p$
\item $c^Z_{0,p+1}=\frac{a+b}{2}$, $c^Z_{i,p+1}=0$ for all 
$i=1,\ldots,n$ and $c^Z_{n+1,p+1}=\frac{\mid a-b \mid}{2}$
\item $p^Z_{j,k}=p^X_{j,k}$ for all $j=1,\ldots,m$, $k=1,\ldots, p$
\item $p^Z_{j,p+1}=0$ for all $j=1,\ldots,m$
\end{itemize}
Or in block matrix form, 
$C^Z=\left(\begin{array}{c|c}
& \frac{a+b}{2} \\
& 0 \\
C^X 
& \ldots \\
& 0\\
\hline
0 & \frac{\mid a-b \mid}{2}
\end{array}\right)$, 
$P^Z=\left(\begin{array}{c|c}
& 0 \\
P^X & \ldots \\
& 0 \\
\end{array}\right)$
\end{definition}

We carry on by addition, or more precisely, the operation interpreting
the assignment $x_{p+1}:=x_i+x_j$ and adding new variable $x_{p+1}$ to the affine set:

\begin{definition}
Let $X=(C^X,P^X)$ be a \pform $\,$  in ${\cal M}(n+1,p) \times {\cal M}(m,p)$. 
We define $Z=\semb x_{p+1}=x_i + x_j \seme X=(C^Z,P^Z)\in {\cal M}(n+1,p+1) \times {\cal M}(m,p+1)$ by \\
$$C^Z=\left(\begin{array}{c|c}
C^X & \begin{array}{c}
c^X_{0,i}+c^X_{0,j} \\
\ldots \\
c^X_{n,i}+c^X_{n,j} \\
\end{array}
\end{array}\right) \mbox{ and } \;
P^Z=\left(\begin{array}{c|c}
P^X & \begin{array}{c}
p^X_{1,i}+p^X_{1,j} \\
\ldots \\
p^X_{m,i}+p^X_{m,j} \\
\end{array}
\end{array}\right).$$
\end{definition}
Finally, we give a meaning to the interpretation of assignments of
the form $x_{p+1}:=\lambda x_i$, for $\lambda \in \R$~:

\begin{definition}
Let $X=(C^X,P^X)$ be a \pform $\,$ in ${\cal M}(n+1,p) \times {\cal M}(m,p)$. 
We define $Z=\semb x_{p+1}=\lambda x_i \seme X=(C^Z,P^Z)\in {\cal M}(n+1,p+1) \times {\cal M}(m,p+1)$ by \\
$$C^Z=\left(\begin{array}{c|c}
C^X & \begin{array}{c}
\lambda c^X_{0,i}\\
\ldots \\
\lambda c^X_{n,i}\\
\end{array}
\end{array}\right) \mbox{ and } \;
P^Z=\left(\begin{array}{c|c}
P^X & \begin{array}{c}
\lambda p^X_{1,i}\\
\ldots \\
\lambda p^X_{m,i}\\
\end{array}
\end{array}\right).$$
\end{definition}
We can prove the correctness of our abstract semantics:

\begin{lemma}\label{arithaffine}
Operations $X \rightarrow \semb x_{p+1}=[a,b] \seme X$, $X \rightarrow \semb x_{p+1}=x_i + x_j \seme X$ and 
$X \rightarrow \semb x_{p+1}=\lambda x_i \seme X$ are increasing over \pforms.
Moreover 
these three operations
do not introduce over-approximations.  
\end{lemma}

\noindent {\sc Proof.}
Suppose we are given two \pforms $\,$ $X$ and $Y$ such that $X \leq Y$.

First, for constant assignments, we have, for all $t \in \R^{p+1}$:
$$\begin{array}{rcl}
\norm{(C^{\semb x_{p+1}=[a,b] \seme X}-C^{\semb x_{p+1}=[a,b] \seme Y})t} & = & \norm{(C^X-C^Y)t} \\
& \leq & \norm{P^Y t} - \norm{P^X t} \\
& \leq & \norm{P^{\semb x_{p+1}=[a,b] \seme Y} t} - \norm{P^{\semb x_{p+1}=[a,b] \seme X} t}
\end{array}$$
which shows monotonicity of $X \rightarrow \semb x_{p+1}=[a,b] \seme X$
The concretisation of $\semb x_{p+1}=[a,b] \seme X$ is obviously exact.

\def\norm#1{\mbox{$\| #1 \|_1$}}

Now for addition of variables, we have, for all $t \in \R^{p+1}$:\\
$$\begin{array}{rcl}
\| (C^{\semb x_{p+1}=x_i + x_j \seme X}&-&C^{\semb x_{p+1}=x_i + x_j \seme Y})t \|_1  = \\ &= &
\sum_{l=0}^n \mid \sum_{k=0}^{p+1} 
(c^{\semb x_{p+1}=x_i + x_j \seme X}_{l,k}-c^{\semb x_{p+1}=x_i + x_j \seme Y}_{l,k}) t_k \mid\\
& = & \sum_{l=0}^n \mid \sum_{k=0}^{p} 
(c^{X}_{l,k}-c^{Y}_{l,k}) t_k +
(c^X_{i,k}+c^X_{j,k}) t_{p+1} \mid\\
& = & \norm{(C^X-C^Y) 
\transpose{
(t_1,
\ldots,
t_i+t_{p+1},
\ldots,
t_j+t_{p+1},
\ldots,
t_p)}} \\
& \leq & \norm{P^Y \transpose{
(t_1,
\ldots,
t_i+t_{p+1},
\ldots,
t_j+t_{p+1},
\ldots,
t_p)}}\\
& & -\norm{P^X 
\transpose{
(t_1,
\ldots,
t_i+t_{p+1},
\ldots,
t_j+t_{p+1},
\ldots,
t_p)}} \\
& = & \norm{P^{\semb x_{p+1}=x_i + x_j \seme Y} t}-\norm{P^{\semb x_{p+1}=x_i + x_j \seme X} t}
\end{array}$$
which shows monotonicity of $X \rightarrow \semb x_{p+1}=x_i + x_j \seme X$
The concretisation of $\semb x_{p+1}=x_i + x_j \seme X$ is obviously exact.

And finally, we have, for all $t \in \R^{p+1}$:
$$\begin{array}{rcl}
\| (C^{\semb x_{p+1}=\lambda x_i \seme X}&-&C^{\semb x_{p+1}=\lambda x_i \seme Y})t \|_1  = \\ &=&
\sum_{l=0}^n \mid \sum_{k=0}^{p+1} 
(c^{\semb x_{p+1}=\lambda x_i \seme X}_{l,k}-c^{\semb x_{p+1}=\lambda x_i \seme Y}_{l,k}) t_k \mid\\
& = & \sum_{l=0}^n \mid \sum_{k=0}^{p} 
(c^{X}_{l,k}-c^{Y}_{l,k}) t_k +
\lambda c^X_{i,k} t_{p+1} \mid\\
& = & \norm{(C^X-C^Y) 
\transpose{
(t_1,
\ldots,
t_i+\lambda t_{p+1},
\ldots,
t_p)}} \\
& \leq & \norm{P^Y \transpose{
(t_1,
\ldots,
t_i+\lambda t_{p+1},
\ldots,
t_p)}}\\
& & -\norm{P^X 
\transpose{
(t_1,
\ldots,
t_i+\lambda t_{p+1},
\ldots,
t_p)}} \\
& = & \norm{P^{\semb x_{p+1}=\lambda x_i \seme Y} t}-\norm{P^{\semb x_{p+1}=\lambda x_i \seme X} t}
\end{array}$$
which shows monotonicity of $X \rightarrow \semb x_{p+1}=\lambda x_i \seme X$
The concretisation of $\semb x_{p+1}=\lambda x_i \seme X$ is obviously exact.
\qed

\vskip .5cm

\subsubsection{Polynomial assignments}

\label{mult}

The following operation defines the multiplication of
variables $x_i$ and $x_j$, appending the result to 
the \pform \; $X$. All polynomial assignments can be defined using this and
the previous operations. 

\begin{definition}
Let $X=(C^X,P^X)$ be a \pform $\,$ in ${\cal M}(n+1,p) \times {\cal M}(m,p)$. 
We define $Z=(C^Z,P^Z)=\semb x_{p+1}=x_i \times x_j \seme X \in {\cal M}(n+2,p+1) \times {\cal M}(m+1,p+1)$ by~:
\begin{itemize}
\item $c^z_{i,k}=c^x_{i,k}$ and
$c^z_{n+1,k}=0$ for all $i=0,\ldots,n$ and
$k=1,\ldots,p$
\item $c^z_{0,p+1}=c^x_{0,i} c^y_{0,j}$
\item $c^z_{l,p+1}=c^x_{0,i} c^y_{l,j}+c^x_{l,i} c^y_{0,j}$
for all $l=1,\ldots,n$
\item $c^z_{n+1,p+1}=\sum_{1\leq r, l \leq n} \mid c^x_{r,i} 
c^y_{l,j} \mid$
\item $p^z_{l,k}= p^x_{l,k}$, $p^z_{m+1,k}=0$ and $p^z_{l,p+1}=0$, for
all $l=1,\ldots,m$ and $k=1,\ldots,p$
\item $p^z_{m+1,p+1}=\sum_{1 \leq r, l \leq m} \mid p^x_{r,i}
p^y_{l,j} \mid + \sum_{0 \leq r \leq n}^{1 \leq l \leq m} \mid c^x_{r,i}
p^y_{l,j} \mid + \sum_{0 \leq l \leq n}^{1 \leq r \leq m} \mid p^x_{l,i}
c^y_{r,j} \mid$
\end{itemize}
\end{definition}

\begin{lemma}\label{multincreasing}
The operation $X \rightarrow \semb x_{p+1}=x_i \times x_j \seme X$ is increasing, and
has a concretisation which contains the set of points of the
form $(x_1,\ldots,x_{p+1})$ with $(x_1,\ldots,x_p) \in \gamma(X)$ and 
$x_{p+1}=x_i x_j$.
\end{lemma}

\noindent {\sc Proof.}
Let $X$ and $Y$ be two \pforms \; such that $X \leq Y$, and let $U=\semb x_{p+1}=
x_i \times x_j\seme X$ and $T=\semb x_{p+1}= x_i \times x_j \seme Y$. 
We compute for all $t \in \R^{p+1}$:

$$\begin{array}{rcl}
\norm{(C^T-C^Z)t} & = & \mid \sum_{l=1}^p (c^Y_{0,l}-c^X_{0,l}) t_l +
\left(c^Y_{0,i}c^Y_{0,j}-c^X_{0,i}c^X_{0,j}\right)t_{p+1} \mid \\
& & +\sum_{k=1}^n \mid \sum_{l=1}^p (c^Y_{k,l}-c^X_{k,l})t_l \\
& & +
\left(c^Y_{0,i}c^Y_{k,j}+c^Y_{k,i}c^Y_{0,j}-
c^X_{0,i}c^X_{k,j}-c^X_{k,i}c^X_{0,j}\right) t_{p+1}\mid \\
& & + \mid \sum_{k=1}^n \sum_{l=1}^n \left(\mid c^Y_{k,i} c^Y_{k,j} \mid
- \mid c^X_{k,i} c^X_{k,j} \mid \right) t_{p+1} \mid      \\
& \leq & \mid\sum_{k=0}^n \mid \sum_{l=1}^p (c^Y_{k,l}-c^X_{k,l}) t_l \mid \\
& & +\mid \left((c^Y_{0,i}-c^X_{0,i}) c^Y_{0,j}+c^X_{0,i}(c^Y_{0,j}-c^X_{0,j})\right) t_{p+1}\mid \\
& & +\sum_{k=1}^n \mid (
(c^Y_{k,j}-c^X_{k,j})c^X_{0,i} + c^Y_{k,j}(c^Y_{0,i}-c^X_{0,i})\\
& & +(c^Y_{k,i}-c^X_{k,i})c^Y_{0,j}+ c^X_{k,i}(c^Y_{0,j}-c^X_{0,j})
) t_{p+1} \mid \\
& & + \mid \sum_{k=1}^n \sum_{l=1}^n (
(\mid c^Y_{k,i}\mid - \mid c^X_{k,i}\mid)\mid c^Y_{l,j}\mid\\
& & +\mid c^X_{k,i}\mid (\mid c^Y_{l,j}\mid-\mid c^X_{l,j}\mid)
) t_{p+1} \mid   \\
& \leq & \norm{P^Y t}-\norm{P^X t} \\
& & +\left(\sum_{l=0}^n \mid c^Y_{l,j}\mid \right) \left(\sum_{k=0}^n
\mid c^Y_{k,i}-c^X_{k,i}\mid\right) \mid t_{p+1} \mid \\
& & +\left(\sum_{k=0}^n\mid c^X_{k,i} \mid\right) \left(
\sum_{l=0}^n \mid c^Y_{l,j}-c^X_{l,j}\mid\right) \mid t_{p+1} \mid\\
\end{array}$$
But $X \leq Y$ so $\pi_i(X) \leq \pi_i(Y)$ and $\pi_j(X) \leq \pi_j(Y)$. 
Therefore,
$$\begin{array}{rcl}
\left(\sum_{k=0}^n\mid c^X_{k,i} \mid\right) \left(
\sum_{l=0}^n \mid c^Y_{l,j}-c^X_{l,j}\mid\right) & \leq &
\norm{\pi_i(C^X)}\left(\norm{\pi_j(P^Y)}-\norm{\pi_j(P^X)}\right)
\end{array}$$
and,
$$\begin{array}{rcl}
\left(\sum_{l=0}^n \mid c^Y_{l,j}\mid \right) \left(\sum_{k=0}^n
\mid c^Y_{k,i}-c^X_{k,i}\mid\right) & \leq &
\norm{\pi_j(C^Y)}\left(\norm{\pi_i(P^Y)}-\norm{\pi_i(P^X)}\right)
\end{array}$$
Hence,
$$\begin{array}{rcl}
\norm{(C^T-C^Z)t} & \leq &\norm{P^Y t}
+\norm{\pi_i(C^X)}
\norm{\pi_j(P^Y)}\mid t_{p+1} \mid\\
& & +\norm{\pi_j(C^Y)}
\norm{\pi_i(P^Y)}\mid t_{p+1} \mid \\
& & 
-\norm{P^X t}
-\norm{\pi_i(C^X)}\norm{\pi_j(P^X)} \mid t_{p+1} \mid\\
& & -\norm{\pi_j(C^Y)}\norm{\pi_i(P^X)} \mid t_{p+1} \mid \\
& \leq & \norm{P^Y t}
+\left(\norm{\pi_i(C^X-X^Y)}+\norm{\pi_i(C^Y)}\right)\norm{\pi_j(P^Y)} \mid t_{p+1} \mid \\
& & +\norm{\pi_j(C^Y)}
\norm{\pi_i(P^Y)}\mid t_{p+1} \mid \\
& & -\norm{P^X t}
-\norm{\pi_i(C^X)}\norm{\pi_j(P^X)} \mid t_{p+1} \mid\\
& & \left(\norm{\pi_j(C^X-C^Y)}+\norm{\pi_j(C^X)}\right)\norm{\pi_i(P^X)} \mid t_{p+1} \mid\\
&\leq & \norm{P^Y t}
+(\norm{\pi_i(P^Y)}\norm{\pi_j(P^Y)}+\norm{\pi_i(C^Y)}\norm{\pi_j(P^Y)} \\
& & +\norm{\pi_j(C^Y)}\norm{\pi_i(P^Y)}) \mid t_{p+1} \mid\\
& & -\norm{P^X t}
+(\norm{\pi_i(P^X)}\norm{\pi_j(P^X)}-\norm{\pi_i(C^X)}\norm{\pi_j(P^X)} \\
& & -\norm{\pi_j(C^X)}\norm{\pi_i(P^X)})\mid t_{p+1} \mid
\end{array}$$
Hence the result, since precisely:
$$p^z_{m+1,p+1}=\sum_{1 \leq r, l \leq m} \mid p^x_{r,i}
p^y_{l,j} \mid + \sum_{0 \leq r \leq n, 1 \leq l \leq m} \mid c^x_{r,i}
p^y_{l,j} \mid + \sum_{0 \leq l \leq n, 1 \leq r \leq m} \mid p^x_{l,i}
c^y_{r,j} \mid$$ is also equal to
$$\norm{\pi_i(P^X)}\norm{\pi_j(P^X)}+\norm{\pi_i(C^X)}\norm{\pi_j(P^X)} \\
+\norm{\pi_j(C^X)}\norm{\pi_i(P^X)}$$
Finally, the fact that the image of $x_{p+1}$ contains all the products $x_i\times x_j$ is
trivial.
\qed

\vskip .5cm

\subsection{The join operator}

\label{joinoperator}

We first recall the definition of a minimal upper bound or {\em mub}:

\begin{definition}
Let $\sqsubseteq$ be a partial order on a set $X$. We say that
$z$ is {\em a mub} of two elements $x,y$ of $X$ if and only if
\begin{itemize}
\item $z$ is an upper bound of $x$ and $y$, i.e. $x \sqsubseteq z$ and
$y \sqsubseteq z$,
\item for all $z'$ upper bound of $x$ and $y$,
$z'\sqsubseteq z$ implies $z=z'$.
\end{itemize}
\end{definition}

We give below an example of such mubs on \pforms. 

\begin{example}\label{firstunion}

Consider
$$\begin{array}{ccc}
X = \left(\begin{array}{c}
1+\varepsilon_1 \\
1+\varepsilon_1
\end{array}\right) \;\; & \;\;
Y = \left(\begin{array}{c}
1+2\varepsilon_1 \\
1+2\varepsilon_1
\end{array}\right) 
\;\; & \;\; Z = \left(\begin{array}{c}
1+1.5\varepsilon_1+0.5\eta_1 \\
1+1.5\varepsilon_1+0.5\eta_1
\end{array}\right)
\end{array}$$
$Z$ is a mub for $X$ and $Y$, given by a ``midpoint'' formula.

\end{example}

This gives us an idea on how to 
find, in $O((n+m)p)$ time, a mub in some cases, or a tight upper bound, in all cases:

\begin{lemma}\label{formuleunion}
Let $X=(C^X,P^X)$ and $Y=(C^Y,P^Y)$ be two \pforms \;
in ${\cal M}(n+1,p)\times {\cal M}(m,p)$. Upper bounds $Z=(C^Z,P^Z)$ of
$X$ and $Y$ satisfy:
\begin{equation}
\label{equ8bis}
\forall t \in \R^p, \norm{P^Z t} \geq \frac{1}{2} \left(\norm{(C^Y-C^X)t}
+\norm{P^X t}+\norm{P^Y t}\right)\\
\end{equation}
When $\gamma_{lin}(P^X)=\gamma_{lin}(P^Y)$, there exists a mub $Z$ with $P^Z$ satisfying (\ref{equ8bis})
with equality; it is defined by
$Z=(C^Z,P^Z) \in {\cal M}(n+1, p)\times {\cal M}(m+n+1, p)$ with:
\begin{itemize}
\item $c^Z_{i,k}=\frac{1}{2}\left(c^X_{i,k}+c^Y_{i,k}\right)$ 
for all $i=0,\ldots,n$, $k=1,\ldots,p$
\item $p^Z_{j+1,k}=\frac{1}{2}(c^X_{j,k}-c^Y_{j,k})$ for all
$j=0,\ldots,n$, $k=1,\ldots, p$
\item $p^Z_{n+j+1,k}=p^X_{j,k}$
for all $j=1,\ldots,m$, $k=1,\ldots,p$ 
\end{itemize}
\end{lemma} 

\noindent {\sc Proof.}
We begin by showing the following: 
let $X=(C^X, P^X)$ and $Y=(C^Y,P^Y)$ two \pforms $\,$ in ${\cal M}(n+1,p)\times {\cal M}(m,p)$.
Minimal upper bounds $Z=(C^Z,P^Z)$ of $X$ and $Y$ satisfy:
\begin{equation}
\label{equ8}
\forall t \in \R^p, \norm{P^Z t} \geq \frac{1}{2} \left(\norm{(C^Y-C^X)t}
+\norm{P^X t}+\norm{P^Y t}\right)\\
\end{equation}

As $X \leq Z$ and $Y \leq Z$, we have, for all $t \in \R^p$:
\begin{equation}
\label{XleqZ}
\norm{(C^Z-C^X)t} \leq \norm{P^Z t}-\norm{P^X t} 
\end{equation}
\begin{equation}
\label{YleqZ}
\norm{(C^Z-C^Y)t} \leq \norm{P^Z t}-\norm{P^Y t} 
\end{equation}
So, 
$$\begin{array}{rcl}
\norm{(C^Y-C^X)t} & \leq & \norm{(C^Z-C^Y)t}+\norm{C^Z-C^X)t} \\
& \leq & 2 \norm{P^Z t}-\norm{P^X t}-\norm{P^Y t}
\end{array}$$
Therefore we have inequality \ref{equ8}. 

If ever we find $Z=(C^Z,P^Z)$ such that inequality \ref{equ8} is in fact an
equality, and such that $Z$ is an upper bound of $X$ and $Y$, 
then we are sure that $Z$ is a mub. Since whenever we take another
upper bound $T$ of $X$ and $Y$, $T$ cannot possibly be strictly less
than $Z$, for $\norm{P^Z t}-\norm{P^T t} \leq 0$ by inequality \ref{equ8}.

We 
notice that
the equation on zonotope $P^Z$ given by 
$$\norm{P^Z t}=
\frac{1}{2} \left(\norm{(C^Y-C^X)t}+\norm{P^Y t}+\norm{P^X t}\right)$$
trivially realizing inequality \ref{equ8} as an equality, 
can easily be solved by taking $PZ$ as the Minkowski sum of 
zonotopes given by $C^Y-C^X$, $P^Y$ and
$P^X$ reduced in size by half. An easy choice is to make:
$$P^Z=\frac{1}{2} \left(\begin{array}{c}
C^Y -C^X \\
\hline
P^X \\
\hline
P^Y 
\end{array}\right)$$
or any choice (with less noise symbols for instance) giving the same
zonotope, geometrically.

Now we have found a potential $P^Z$, we rewrite inequalities \ref{XleqZ}
and \ref{YleqZ}:
\begin{equation}
\label{in1}
\norm{(C^Z-C^X)t} \leq \frac{1}{2}\left(\norm{(C^Y-C^X)t}+\norm{P^Y t}
-\norm{P^X t}\right) 
\end{equation}
\begin{equation}
\label{in2}
\norm{(C^Z-C^Y)t} \leq \frac{1}{2}\left(\norm{(C^Y-C^X)t}+\norm{P^X t}
-\norm{P^Y t}\right) 
\end{equation}
In case $\gamma_{lin}(P^X)=\gamma_{lin}(P^Y)$, inequalities \ref{in1} and
\ref{in2} can be made into equalities, choosing $C^Z$ to have
entries being the mean of the corresponding entries of $C^X$ and
$C^Y$, exactly realizing $\norm{(C^Z-C^X)t}=\frac{1}{2}\norm{(C^Y-C^X)t}=
\norm{(C^Z-C^Y)t}$. In that case, we can choose for example 
$$P^Z= \left(\begin{array}{c}
\frac{1}{2} (C^Y -C^X) \\
\hline P^X \end{array}\right).$$
\qed

\vskip .5cm

We do not fully discuss here the general case, but some intuition is given in Example \ref{discmubs}.
A good over-approximation of a mub is given by the above formula applied
to $X'=(C^X, P^U)$ and $Y'=(C^Y, P^U)$, where $P^U$ is such that 
$\gamma(P^X)\cup
\gamma(P^Y) \subseteq \gamma(P^U)$. 

\begin{example}\label{secondunion}
Consider now:
$$\begin{array}{cc}
X = \left(\begin{array}{c}
1+2\varepsilon_1 \\
-1+\varepsilon_1-2\varepsilon_2
\end{array}\right) &
Y = \left(\begin{array}{c}
3+\varepsilon_1 \\
1+2\varepsilon_1-\varepsilon_2
\end{array}\right)
\end{array}$$
Using Lemma \ref{formuleunion}, we find
\[ Z = \left(\begin{array}{c} 2 +1.5\varepsilon_1+\eta_1-0.5\eta_2 \\
1.5\varepsilon_1-1.5\varepsilon_2+\eta_1+0.5\eta_2+0.5\eta_3
\end{array}\right)\]
which is a mub indeed. It is depicted in Figure \ref{dessin2}.
\end{example}

\subsubsection{Convergence acceleration}

The trouble with Lemma \ref{formuleunion} is that it may produce a lot
of new noise symbols, thus being not always easily applicable. We thus introduce a less 
refined join operator, which also very often allows to accelerate fixpoint convergence.
For any interval $i$, we note $mid(i)$ its center. Let $\alpha \wedge \beta$ denote the minimum of the 
two real numbers, and $\alpha \vee \beta$ their maximum. We define
\[
\argmin{\alpha}{\beta} = \{\gamma \in [\alpha \wedge \beta,\alpha \vee \beta],
 |\gamma| \mbox{ minimal}\}
\]

\begin{lemma}\label{nonoptimaljoin}
Let $X=(C^X,P^X)$ and $Y=(C^Y,P^Y)$ be two \pforms \; in ${\cal M}(n+1,p)\times
{\cal M}(m,p)$. We define
$Z=(C^Z,P^Z)=X\nabla Y\in {\cal M}(n+1, p) \times {\cal M}(m+p, p)$ by:
\begin{itemize}
\item $c^Z_{0,k}=mid\left(\gamma(\pi_k(X))\cup\gamma(\pi_k(Y))\right)$
for all $k=1,\ldots,p$
\item $c^Z_{i,k}=\argmin{c^X_{i,k}}{c^Y_{i,k}}$
for all $i=1,\ldots,n$, $k=1,\ldots,p$
\item $p^Z_{j,k}=\argmin{p^X_{j,k}}{p^Y_{j,k}}$ for all
$j=1,\ldots,m$, $k=1,\ldots, p$
\item $p^Z_{m+j,j}=
\sup \gamma(\pi_j(X))\cup\gamma(\pi_j(Y))
-\sup \gamma\left(c^Z_{0,j}+\sum_{i=1}^n c^Z_{i,j} \varepsilon_j
+\sum_{i=1}^m p^Z_{i,j} \eta_j\right)$ for all $j=1,\ldots,p$
\item $p^Z_{m+j,k}=0$ for all $j, k=1,\ldots,p$ with $j\neq k$
\end{itemize}
Then $Z$ is an upper bound of $X$ and $Y$ such that
for all $k=1,\ldots,p$, $\gamma(\pi_k(Z))=\gamma(\pi_k(X))\cup 
\gamma(\pi_k(Y))$
\end{lemma}

\noindent {\sc Proof.}
We prove that $X \leq Z$, the property that $\gamma(\pi_k(Z))=
\gamma(\pi_k(X))\cup\gamma(\pi_k(Y))$ being easy to check (by construction!). 
Now, we want to prove negativity, for all $t \in \R^p$
of:
$$\sum_{i=0}^n \mid \sum_{k=1}^p (c^Z_{i,k}-c^X_{i,k})t_k \mid
+\sum_{j=1}^m\mid \sum_{k=1}^p p^X_{j,k} t_k \mid
-\sum_{j=1}^{m} \mid \sum_{k=1}^p p^Z_{j,k} t_k \mid
-\sum_{j=1}^p \mid p^Z_{m+j,j} t_j \mid$$
By the triangular inequality, the sum of the first 2 terms is less or equal to
$$\sum_{i=0}^n \mid \sum_{k=1}^p (c^Z_{i,k}-c^X_{i,k})t_k \mid
+\sum_{j=1}^m\mid \sum_{k=1}^p (p^Z_{j,k}-p^X_{j,k}) t_k \mid$$
then using it again for each sum, is less or equal to
$$\sum_{k=1}^p \mid t_k \mid 
\left(\sum_{i=1}^n \mid c^Z_{i,k}-c^X_{i,k} \mid 
+\sum_{j=1}^m\mid p^Z_{j,k}-p^X_{j,k} \mid\right)$$
But we know by \cite{DBLP:journals/corr/abs-0807-2961}, section 3.5.1, where this operator for accelation of convergence was defined, 
that for all $k=1,\ldots,p$,
$\sum_{i=0}^n\mid c^Z_{i,k}-c^X_{i,k} \mid +
\sum_{j=1}^m \mid p^Z_{j,k}-p^X_{j,k}\mid
\leq \mid p^Z_{m+k,k} \mid$.
So overall, this is less than $\sum_{k=1}^p \mid p^Z_{m+k,k} t_k \mid$.
\qed

\vskip .5cm

This $\nabla$ operation may be sub-optimal, but the concretisations on each
axis (i.e. the immediate concretisation of all program variables) are optimal. 
Also, while its cost of computation is still of $O((n+m)p)$, it may produce far
less perturbation symbols, and may even kill over some of the central symbols.

\begin{figure}
\begin{center}
\begin{tikzpicture}[xscale=1.2,yscale=0.4] 
\filldraw[fill=green!20!white,draw=green!50!black,dashed,fill opacity=0.5] (-1,-4) -- (0,-5) -- (5,0) -- (5,4) -- (4,5) -- (-1,0) -- (-1,-4);
\coordinate [label=above:\textcolor{green!50!black}{$Z$}] (Z) at (0.15,-4.9);
\filldraw[fill=red!20!white,draw=red!50!black,dashed,fill opacity=0.5] (-1,-4) -- (2,-4) -- (4,-2) -- (4,4) -- (1,4) -- (-1,2) -- (-1,-4);
\coordinate [label=below:\textcolor{red!50!black}{$Z'$}] (Z1) at (-0.85,2.);
\filldraw[fill=blue!20!white,draw=blue!50!black,dashed,fill opacity=0.5] (-1,-4) -- (1,-4) -- (3,-2) -- (4,0) -- (4,4) -- (2,4) -- (0,2) -- (-1,0) -- (-1,-4);
\coordinate [label=below:\textcolor{blue!50!black}{$Z"$}] (Z2) at (2.15,4.);
\draw[fill=yellow!50!white,draw=black,fill opacity=0.5] (-1,-4) -- (3,-2) -- (3,2) -- (-1,0) -- (-1,-4); 
\coordinate [label=above:\textcolor{black}{$X$}] (X) at (-0.9,-3.9);
\filldraw[fill=yellow!50!white,draw=black,fill opacity=0.5] (2,-2) -- (4,2) -- (4,4) -- (2,0) -- (2,-2);
\coordinate [label=above:\textcolor{black}{$Y$}] (Y) at (2.1,-1.9);
\coordinate [label=above:\textcolor{black}{$X$}] (X) at (-0.9,-3.9);
\draw[step=1.cm,gray,very thin] (-2,-5) grid (5,5); 
\draw[->] (-2,0) -- (5,0) node[right]{$x$} ;
\draw[->] (0,-5) -- (0,5) node[right]{$y$};
\foreach \x/\xtext in {-2,-1,...,5} 
\draw (\x cm,0 cm)  node[below] {$\xtext$};
\foreach \y/\ytext in {-5,-4,...,5} 
\draw (0 cm,\y cm) node[left] {$\ytext$};
\end{tikzpicture} 
\caption{$Z$ and $Z''$ are mubs for $X$ and $Y$, while $Z'$ is not \label{dessin2}}
\end{center}
\vspace{-0.5cm}
\end{figure}
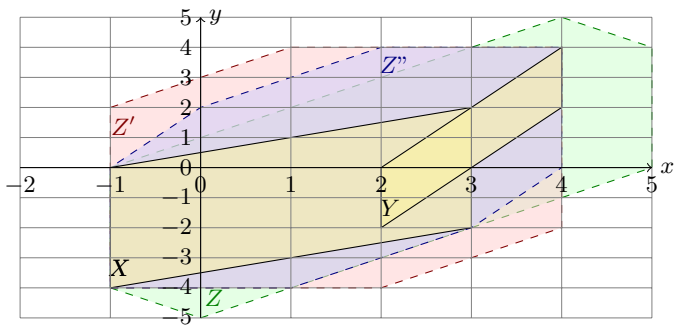

\begin{example}
\label{discmubs}
Consider X and Y as defined in Example \ref{secondunion}:
$$Z' = X \nabla Y = \left(\begin{array}{c}
1.5 +\varepsilon_1+1.5\eta_1 \\
\varepsilon_1-\varepsilon_2+2\eta_2
\end{array}\right)$$
Note that (see Figure \ref{dessin2}) $Z'$ has the smallest possible concretisations
on the $x$ and $y$ coordinates: respectively $[-1,4]$ and $[-4,4]$,  which is
strictly better than what we had with the mub $Z$ in Example \ref{secondunion} 
(respectively $[-1,5]$ and $[-5,5]$). But it does not share perturbation noise symbols, as $Z$ does, 
and along direction
$t=\transpose{(-1,1)}$, we find $Z't=y-x \in [-6,3]$ which
is not as good as we had with $Z$: $Z t \in [-5,1]$. 
In fact, $Z$ and $Z'$ are not comparable under $\leq$. But $Z'$ is not
a mub, just consider:  
$$Z'' = \left(\begin{array}{c}
1.5 +\varepsilon_1+0.5\eta_1+\eta_2 \\
\varepsilon_1-\varepsilon_2+\eta_1+ \eta_3
\end{array}\right)$$
We can prove that $Z'' \leq Z'$, and in fact, $Z''$ is a mub.
$Z''$ has the smallest possible concretisations on the $x$ and $y$ axes as shown in Figure \ref{dessin2}, 
but $Z''t \in [-5,2]$ which is not as accurate as $Z t$ : $Z$ and $Z''$ are also incomparable. 
\end{example}

\subsection{Kleene-like iteration schemes}

\label{completeness}

We first note that we have enough mubs so that to hope for
a Kleene-like iteration:

\begin{lemma}\label{quasicomplete}
Let $S$ be a bounded and countable directed set of \pforms \;
all in ${\cal M}(n+1,p) \times {\cal M}(m,p)$. 
Then there exists a minimal upper bound
for $S$, given by the limit matrices $\stackreh{lim}{u\rightarrow \infty} X_u=
(\stackreh{lim}{u\rightarrow \infty}
C^{u},\stackreh{lim}{u\rightarrow \infty} P^{u})$.
\end{lemma}

\noindent {\sc Proof.}
We thus have $X$ a \pform \; and $$S=\{X_0,\ldots, X_u,\ldots\}$$ with 
$X_i \leq X_j\leq x$ for all $i$, $j$ with $i\leq j$. 
Thus for all $t \in \R^p$, $$\norm{(C^j-C^i)t} \leq \norm{P^jt}-\norm{P^it}$$

This entails
first that $(\norm{P^ut})_{u\in \N}$ is increasing. Also, as for all $u$, $X_u \leq X$, this means
that $0 \leq \norm{(C^X-C^u)t}\leq \norm{P^Xt}-\norm{P^u t}$, so the
sequence $(\norm{P^ut})_{u\in \N}$ is also bounded by
$\norm{P^X t}$. Hence it is converging for all $t$. 

This means also that $\norm{(C^j-C^i)t}$ can be made
as small as wanted with $i$ and $j$ sufficiently big, for all $t$. Hence, as $(\R^p,\norm{.})$ is
a Banach space, this means that for all $t$, $C^u t$ converges when $u$ goes to the infinity.
This entails the convergence of the sequence of matrices $C^u$ in the
{\em fixed dimension} space ${\cal M}(n+1,p)$, similarly for
$P^u$ in ${\cal M}(m,p)$. 

Note that this finite dimension requirement is
necessary. As for polyhedra, an infinite union of zonotopes might not be
a zonotope: just think of a zonotope with a growing number of faces, 
approximating a circle.    

The fact that the limit matrices define a minimal upper bound is an obvious
consequence of the fact that the order $\leq$ is closed in 
$\left({\cal M}(n+1,p)
\times {\cal M}(m,p)\right)^2$, and of basic properties of limits.
 \qed

\vskip .5cm

As we have only this form of bounded completeness, and not inconditional completeness,
our iteration schemes will be parameterized by a large interval $I$: as soon
as the current iterate leaves $I^p$, we end iteration by $\top$.

The following formalizes the iteration scheme and stopping criterion used, parametrized
by a join operator (for instance, the $\nabla$ operator defined in Lemma \ref{nonoptimaljoin}):

\begin{definition}
\label{iter}
Given an upper-bound operator $U$, 
the $U$-iteration scheme for a strict, continuous and increasing
functional $F$ on \pforms \; (extended with a formal $\bot$ and $\top$),  
is as follows:
\begin{itemize}
\item Start with $X_0=\bot$
\item Then iterate: $X_{u+1}=X_u U F(X_u)$ starting with $u=1$
\begin{itemize}
\item if $\gamma(X_{u+1}) \subseteq \gamma(X_u)$ then stop with $X_{u+1}$
\item if $\gamma(X_{u+1})\not \subseteq I^p$, then end with $\top$
\end{itemize}
\end{itemize}
\end{definition}

Note that our semantic operators only produce continuous and increasing functionals
$F$. 
Also, initial and cyclic unfoldings are generally applied on top of this iteration 
scheme, so as to
improve the precision of the analysis, see \cite{CAVT09,DBLP:journals/corr/abs-0807-2961}, and we cut the iteration after a finite time. We prove below
the correctness of this scheme and of its stopping criterion. We also indicate its worst-case complexity:

\begin{lemma}\label{stoppingcriterion}
Let $F$ be a strict, continuous and increasing 
functional on \pforms.
Consider the $U$-iteration scheme of Definition
\ref{iter}. Then $\gamma(X_{u+1})\subseteq \gamma(X_u)$ can be checked in 
$O(p (n+m)^2)$ time, and guarantees that $X_{u+1}$ is a post-fixed point of $F$.
\end{lemma}

\noindent {\sc Proof.}
We consider the countable and directed set $S=\{X_u \mid u \in \N\}$
where $X_u=U_{j=0}^u F^j(\bot)$. If it is unbounded, the $U$-iteration scheme will end up with $\top$ in a finite time. Otherwise, apply Lemma \ref{quasicomplete}. Define
$G=F U Id$; it is continuous and 
$G(\stackreh{lim}{u\rightarrow \infty} X_u)=
\stackreh{lim}{u\rightarrow \infty} G(X_u)=
\stackreh{lim}{u\rightarrow \infty} X_u$, so the limit of the $U$-iteration
scheme is a fixed-point of $G$, i.e. a post-fixed point of $F$. The
test $\gamma(X_{u+1}) \subset \gamma(X_u)$, given that $X_u \leq X_{u+1}$
of course, is enough for checking if we reached the limit.
We have already proven that if the stopping criterion is correct, then
the $U$-iteration scheme converges towards $\top$ or towards a 
post-fixed point of $F$, in practise in finite time, since we always
cut the iteration scheme after a fixed number of iterations.

Suppose we apply our stopping criterion, i.e. 
$\gamma(X_{u+1}) \subseteq \gamma(X_u)$. But we have
also $X_u \leq X_{u+1}$. Then for all $t \in \R^p$,
$$\begin{array}{lcl}
\norm{C^{X_{u+1}} t} - \norm{C^{X_u} t} & \leq & \norm{P^{X_{u+1}} t} -
\norm{P^{X_{u}} t} \\
\norm{(C^{X_{u+1}}- C^{X_u}) t} & \leq & \norm{P^{X_{u+1}} t} -
\norm{P^{X_{u}} t} 
\end{array}$$
Adding these two inequalities together, we find:
$$ \norm{C^{X_{u+1}} t}+\norm{(C^{X_{u+1}}- C^{X_u}) t}\leq \norm{C^{X_u} t}$$
But the triangular inequality also shows the inverse inequality, therefore:
$$ \norm{C^{X_{u+1}} t}+\norm{(C^{X_{u+1}}- C^{X_u}) t}= \norm{C^{X_u} t}$$
So we have also:
$$\norm{(C^{X_{u+1}}- C^{X_u}) t} \geq \norm{P^{X_{u}} t} -
\norm{P^{X_{u+1}} t}$$
This implies that for all $t \in \R^p$, 
$\norm{(C^{X_{u+1}}- C^{X_u}) t}=0$ and $\norm{P^{X_{u}} t}=\norm{P^{X_{u+1}} t}$, i.e.
$X_{u+1} \sim X_u$. Hence this implies that if we stop using this criterion, 
then we
stop at a postfixed point of $F$. 
\qed

\vskip .5cm

\ \\
In practice, we use the simpler $O((n+m)p)$ time test: $\forall k=1,\ldots,p$, 
$\norm{X_{u+1} t_k} \leq \norm{X_u t_k}$ first, where $t_k$ is the vector with all 0 entries,
except at position $k$. It is only when this test is true that we compute the
full test $\gamma(X_{u+1}) \subseteq \gamma(X_u)$. \\

Results on fixed-point computations, and comparisons with other abstract 
domains such as polyhedra, are described for preliminary versions of this domain 
in \cite{CAVT09,DBLP:journals/corr/abs-0807-2961}. 
We plan to develop them for this domain in a longer version. 

\section{Conclusion and future work}

We set up a formal framework for a fast and accurate abstract analysis
based on zonotopes. There are several directions from there. 
First of all, we did not thoroughly detail the best way to compute (minimal) upper bounds, 
this will be done in the longer version. 

Secondly, as can be noticed with the analysis of function \texttt{f}
of Section \ref{introex}, the perturbation symbol $\eta_1$ can
be associated with the \texttt{if} statement, with discrete values
$\{-1,1\}$ expressing whether the control flow went through the true or
the false branch. This can be generalized to encode some of the 
interesting (semantical) disjunctive information, necessary for reaching
precise invariants.

Third, a drawback of our domain is that tests are in general not interpreted. 
We are currently thinking of a simple and elegant extension, that
would allow for computing accurate intersections.

Last but not least, we plan to carry on the study initiated in 
\cite{DBLP:journals/corr/abs-0807-2961}. Given a program implementing
a concrete numerical scheme, our abstraction gives us a perturbed 
numerical scheme, that can be studied for convergence similarly to 
the concrete scheme. We started with linear recursive filters where we
had very good results, but this is likely to extend to some non-linear
iterative schemes of wide interest.

\bibliographystyle{plain}
\bibliography{FA2010_arxiv}

\end{document}